\documentclass[a4paper]{jpconf}
\usepackage{graphicx}
\begin{document}
\title{Current noise and Coulomb effects in superconducting contacts}

\author{Artem V. Galaktionov$^1$ and Andrei D. Zaikin $^{2,1}$}

\address{$^1$I.E. Tamm Department of
Theoretical Physics, P.N. Lebedev Physics Institute, 119991
Moscow, Russia\\
$^2$ Institut f\"ur Nanotechnologie,
Karlsruher Institut f\"ur Technologie (KIT), 76021 Karlsruhe, Germany}

\ead{andrei.zaikin@kit.edu}

\begin{abstract}
We derive an effective action for contacts between superconducting terminals with arbitrary transmission distribution of conducting channels. In the case of normal-superconducting (NS) contacts we evaluate interaction correction to Andreev
conductance and demonstrate a close relation between Coulomb
effects and shot noise in these systems. In the case of superconducting (SS) contacts we derive the electron-electron
interaction correction to the Josephson current. At $T=0$ both corrections are found to vanish for fully transparent NS and SS contacts indicating the absence of Coulomb effects in this limit.
\end{abstract}

\section{Introduction}

Low energy electron transport across the
interface between normal metals and superconductors (NS) is
provided by the mechanism of Andreev reflection \cite{And}. This
mechanism involves conversion of a subgap quasiparticle entering
the superconductor from the normal metal into a Cooper pair
together with simultaneous creation of a hole that goes back into
the normal metal. Each such act of electron-hole reflection
corresponds to transferring twice the electron charge $e^*=2e$
across the NS interface and results, e.g., in non-zero conductance of the system at subgap energies \cite{BTK}.

Andreev reflection is also responsible for dc Josephson effect in superconducting weak links without tunnel barriers. Suffering Andreev reflections at both $NS$ interfaces, quasiparticles with energies below the superconducting gap are effectively ``trapped'' inside the junction forming a discrete set of levels which can be tuned by passing the supercurrent across the system \cite{SNSK}. At the same time, these subgap Andreev levels themselves contribute to the supercurrent  \cite{SNSK,SNSI,SNSB,SNSGZ} thus making the behavior of superconducting point contacts and $SNS$ junctions in many respects different from that of tunnel barriers.

Note that the above theories remain applicable if one can neglect Coulomb effects. In small-size superconducting contacts, however, such effects can be
important and should in general be taken into account. A lot is known about interplay between fluctuations
and charging effects in superconducting tunnel barriers \cite{SZ}. Here we examine the properties of superconducting junctions going beyond the tunneling limit. Below we will demonstrate that Coulomb blockade in such junctions weakens with increasing barrier transmissions and eventually disappears in the limit of fully open superconducting contacts. We will also argue that in superconducting systems -- similarly to normal contacts \cite{GZ01} -- there also exists a direct relation between Coulomb effects and current fluctuations.

\section{The model and effective action}
As it is shown in Fig. 1, we will consider
big metallic reservoirs one of which is superconducting while another one could be either normal or superconducting. These two reservoirs are connected by a rather
short normal bridge (conductor) with arbitrary transmission
distribution $T_n$ of its conducting modes
and normal state conductance $G_N\equiv 1/R_N=(e^2/\pi)\sum_n T_n$. Both phase and energy relaxation of electrons may occur only in the
reservoirs and not inside the conductor which length is assumed to
be shorter than dephasing and inelastic relaxation lengths. As usually, Coulomb interaction between electrons in the
conductor area is accounted for by some effective capacitance $C$.

\begin{figure}
\includegraphics[width=6.8cm]{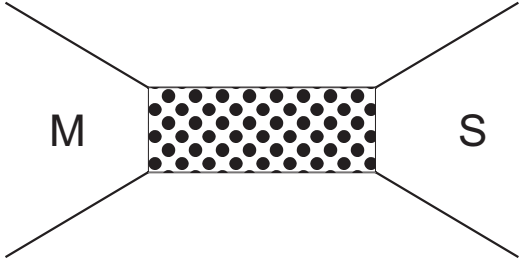}
\caption{The system under consideration. Two big metallic reservoirs -- one superconducting (S) and another one either normal or superconducting (M=N,S) -- are connected by a short normal conductor.}
\end{figure}

In order to analyze electron transport in the presence of
interactions we will make use of an approach based on
the effective action formalism
combined with the scattering matrix technique \cite{GZ01,GGZ03,KN,GZ04,BN}. This approach can be conveniently generalized to superconducting systems. In fact, the structure of the effective action remains the same also in the supeconducting case, one should only
replace normal propagators by $2\times2$ matrix Green functions
which account for superconductivity, as it was done, e.g., in
\cite{SZ,Z,SN}.

Following the standard procedure we express the kernel $J$ of the
evolution operator on the Keldysh contour in terms of a path
integral over the fermionic fields which can be integrated out
after the standard Hubbard-Stratonovich decoupling of the
interacting term. Then the kernel $J$ takes the form
\begin{equation}
J=\int {\cal D} \varphi_1{\cal D}\varphi_2\exp(iS[\varphi]),
\label{pathint}
\end{equation}
where $\varphi_{1,2}$ are fluctuating phases defined on the
forward and backward parts of the Keldysh contour and related to
fluctuating voltages $V_{1,2}$ across the conductor as
$\dot\varphi_{1,2}(t)=eV_{1,2}$. Here and below we set $\hbar =1$.

The effective action consists of two terms, $S[\varphi
]=S_c[\varphi ]+S_t[\varphi ]$, where
\begin{eqnarray}
iS_c[V]= \frac{C}{2e^2}\int\limits_0^t dt'
(\dot\varphi_{1}^2-\dot\varphi_{2}^2)\equiv
\frac{C}{e^2}\int\limits_0^t dt
\dot\varphi^+\dot\varphi^-\label{Sc}
\end{eqnarray}
describes charging effects and the term $S_t[V]$ accounts for
electron transfer between normal and superconducting reservoirs.
It reads \cite{SN}
\begin{equation}
S_t[\varphi]=-\frac{i}{2}\sum_n{\rm Tr} \ln \left[
1+\frac{T_n}{4}\left( \left\{ \check G_M, \check G_S
\right\}-2\right) \right],
\label{St}
\end{equation}
where $\check G_M$ and $\check G_S$ are $4\times4$ Green-Keldysh
matrices of M- and S-electrodes which product
implies time convolution and which anticommutator is denoted by
curly brackets. In Eq. (\ref{Sc}) we also introduced ``classical''
and ``quantum'' parts of the phase, respectively
$\varphi_+=(\varphi_1+\varphi_2)/2$ and
$\varphi_-=\varphi_1-\varphi_2$.

For later purposes we also express the average current and the current-current correlator via the effective action as
\begin{eqnarray}
&&\langle \hat I(t)\rangle =ie\int {\cal D} \varphi_{\pm}\frac{\delta}{\delta
  \varphi_-(t)}e^{iS[\varphi]}, \label{curr}\\
&&\frac12\langle \hat I\hat I\rangle_+=-e^2\int {\cal D} \varphi_{\pm}\frac{\delta^2}{\delta
  \varphi_-(t)\delta\varphi_-(t')}e^{iS[\varphi]},
\label{corr}
\end{eqnarray}
where $\langle \hat I\hat I\rangle_+=\langle \hat I(t)\hat
I(t')+\hat I(t')\hat I(t)\rangle$.

Let us introduce the matrix $\check
X_0[\varphi_+]=1-T_n/2+(T_n/4)\left\{ \check G_M, \check G_S
\right\}|_{\varphi_-=0}$. As the action $S_t$ vanishes for
$\varphi_-(t)=0$ one has ${\rm Tr}\ln \check X_0=0$. Making use of
this property we can identically transform the action (\ref{St})
to
\begin{equation}
S_t=-\frac{i}{2}\sum_n{\rm Tr} \ln \left[ 1+\check X_0^{-1}\circ
\check X' \right], \label{xx}
\end{equation}
where $\check X'=1+(T_n/4)\left( \left\{ \check G_M, \check G_S
\right\}-2\right)-\check X_0$. Now let us separately consider NS and
SS interfaces.

\subsection{NS interfaces}

At temperatures and voltages well
below the superconducting gap Andreev contribution to the action
of NS system dominates. Hence, it suffices to consider the limit of low energies $\epsilon \ll \Delta$. In this limit we can define the Andreev transmissions \cite{BTK} ${\cal T}_n=T^2_n/(2-T_n)^2$  and Andreev conductance $G_A=(2e^2/\pi )\sum_n{\cal T}_n$. Let us assume that either dimensionless Andreev conductance
$g_A=4\sum_n{\cal T}_n$ is large, $g_A \gg 1$, or temperature is
sufficiently high (though still smaller than $\Delta$). In either
case one can describe quantum dynamics of the phase variable
$\varphi$ within the quasiclassical approximation
\cite{GZ01,GGZ03} which amounts to expanding $S_t$ in powers of
(small) ``quantum'' part of the phase $\varphi_-(t)$. Employing
the above equations and expanding $S_t$ up to terms $\sim
\varphi_-^2$ we arrive at the Andreev effective action \cite{GZ09}
\begin{equation}
iS_t=iS_R-S_I, \label{finalS}
\end{equation}
where
\begin{eqnarray}
iS_R&=&-\frac{ig_A}{2\pi}\;\int\limits_0^t dt'\;
\varphi^-(t')\dot\varphi^+(t'), \label{SRNS}\\
S_I&=&-\frac{g_A}{4}\int\limits_{0}^{t}dt'\int\limits_{0}^{t}dt''
\frac{T^2}{\sinh^2[\pi T(t'-t'')]} \varphi^-(t')\varphi^-(t'')
\nonumber\\
&& \times [\beta_A \cos(2\varphi^+(t')-2\varphi^+(t'')) +1-\beta_A
] \label{SINS}
\end{eqnarray}
and
\begin{equation}
\beta_A=\frac{\sum_n{\cal T}_n(1-{\cal T}_n)}{\sum_n{\cal T}_n}
\end{equation}
is the Andreev Fano factor defined in a complete analogy with the
normal Fano factor $\beta_N=\sum_nT_n(1-T_n)/\sum_nT_n$.
We observe that the action $S_t$ is expressed in
exactly the same form as that for normal conductors
\cite{GZ01,GGZ03} derived within the the same quasiclassical
approximation for the phase variable $\varphi (t)$. In order to
observe the correspondence between the action
\cite{GZ01,GGZ03} and that defined in Eqs.
(\ref{finalS})-(\ref{SINS}) one only needs to interchange normal and
Andreev conductances as well as the corresponding Fano factors
\begin{equation}
G_N\leftrightarrow G_A, \quad \beta_N\leftrightarrow \beta_A
\label{transform}
\end{equation}
and to account for an extra factor 2 in front of the
phase $\varphi_+$ under $\cos$ in Eq. (\ref{SINS}). This extra
factor implies doubling of the charge during Andreev reflection.

\subsection{Superconducting contacts}

Turning to superconducting contacts we assume that fluctuating phases $\varphi_\pm (t)$ are sufficiently small
and perform regular expansion of the exact effective action in powers of these phases. Then we obtain
\begin{equation}
iS_t=-\frac{i}{e}\int\limits_0^t dt'I_S(\chi)\varphi_-(t')+ iS_R-S_I, \label{finalsS}
\end{equation}
where $\chi$ is the time-independent phase difference,
\begin{eqnarray}
I_S(\chi )=\frac{e\Delta\sin\chi}{2}\sum_n\frac{T_n}{\sqrt{1-T_n\sin^2(\chi/2)}}
\tanh\frac{\Delta\sqrt{1-T_n\sin^2(\chi/2)}}{2T}.
\label{Ichi}
\end{eqnarray}
defines the supercurrent across the system \cite{KO,SNSB} and
\begin{eqnarray}
S_R&=&\int\limits_0^t dt'\int\limits_{0}^{t}dt''
{\cal R}(t'-t'') \varphi^-(t')\varphi^+(t''), \label{SRRR}\\
S_I&=&\int\limits_{0}^{t}dt'\int\limits_{0}^{t}dt''
{\cal I}(t'-t'') \varphi^-(t')\varphi^-(t'') \label{SIII}
\end{eqnarray}
with both kernels ${\cal R}(t)$ and ${\cal I}(t)$ being real functions. The complete expressions for these functions turn out to be somewhat lengthy and for this reason are not presented here. Below we only emphasize some of the properties of ${\cal R}(t)$ and ${\cal I}(t)$.

To begin with, it is straightforward to verify that in the lowest order in barrier
transmissions $T_n$ the result (\ref{finalsS})-(\ref{SIII}) reduces to the standard AES action \cite{SZ} for tunnel barriers
in the limit of small phase fluctuations. Qualitatively new features emerge in higher orders in $T_n$ being directly related to the presence of subgap Andreev levels $\pm\epsilon_n(\chi)$ inside the contact. Consider, for instance, the kernel ${\cal I}(t)$. It can be split into three contributions of  different physical origin
\begin{equation}
 {\cal I}(t)={\cal I}_1(t)+{\cal I}_2(t)+{\cal I}_3(t).
 \end{equation}
The first of these terms, ${\cal I}_1(t)$, represents the subgap contribution due to discrete Andreev states. The Fourier transform of this term has the form
\begin{eqnarray}
&&{\cal I}_{1\omega}=\frac{\pi\Delta^4}{4}\sum_n \bigg\{  \frac{ T_n^2\sin^2\chi}{2\epsilon_n^2(\chi)\cosh^2(\epsilon_n(\chi)/2T)}\delta(\omega)
\nonumber\\
&&+\frac{ T_n^2(1-T_n)\sin^4(\chi/2)}{\epsilon^2_n(\chi)}\left[1+\tanh^2(\epsilon_n(\chi)/2T)\right]
[\delta\left(\omega-2\epsilon_n(\chi)\right)+ \delta\left(\omega+2\epsilon_n(\chi)\right)] \bigg\}.
\label{I1}
 \end{eqnarray}
It is obvious that this contribution is not contained in the AES action at all. The second term  ${\cal I}_2(t)$
can be interpreted as the ''interference term'' between
subgap Andreev levels and quasiparticle states above the gap. In the low
temperature limit $T \to 0$ the Fourier transform of this term ${\cal I}_{2\omega}$ differs from zero only at sufficiently high frequencies $|\omega |>\Delta +\epsilon_n(\chi)$. At higher temperatures $T >\epsilon_n(\chi)$, however, ${\cal I}_{2\omega}$ vanishes only for $|\omega |<\Delta -\epsilon_n(\chi )$ and remains non-zero otherwise. In the limit
of small barrier transmissions this term scales as ${\cal I}_2\propto T_n^{3/2}$ and, hence, is not contained in the AES action either.
Finally, the third term ${\cal I}_3(t)$ accounts for the contribution of quasiparticles with energies above the gap. In the high frequency limit $\omega \gg \Delta$ or for
$\Delta \to 0$ this term reduces to the standard result for a normal conductor ${\cal I}_{3\omega}\to (\omega /2e^2R_N)\coth (\omega /2T)$.

Turning now to the function ${\cal R}(t)$ in  Eq. (\ref{SRRR}) we note that its Fourier transform can be represented as
${\cal R}_\omega={\cal R}'_\omega+i {\cal R}''_\omega$, where both ${\cal R}'_\omega$ and  ${\cal R}''_\omega$ are real functions. The function ${\cal R}'_\omega$ is even in $\omega$ while ${\cal R}''_\omega$ is an odd function of $\omega$, thus implying that the function ${\cal R}(t)$ is real.

The functions ${\cal R}(t)$ and ${\cal I}(t)$ are not independent. For instance, the Fourier transform ${\cal R}''_\omega$ is related to ${\cal I}_\omega$ by means of the fluctuation-dissipation relation
${\cal R}''_\omega=2 {\cal I}_\omega \tanh (\omega /2T)$.
The two functions ${\cal R}'_\omega$ and ${\cal R}''_\omega$ are in turn linked to each other by the causality principle: the function ${\cal R}(t)$ should vanish for $t<0$.

Finally we would like to point out that with the aid of the above Gaussian effective action one can
easily evaluate the phase-phase correlation functions for our problem. Combining Eqs. (\ref{finalsS})-(\ref{SIII}) with (\ref{Sc}) one finds (cf., e.g. \cite{GZ2})
\begin{eqnarray}
&& \langle \varphi_+(t_1)\varphi_+(t_2)\rangle=
-\int\limits_{-\infty}^{\infty} \frac{d\omega}{2\pi} {\rm Im}\left(\frac{1}{C\omega^2/e^2+{\cal R}_\omega} \right)\coth \frac{\omega}{2 T}e^{-i\omega(t_1-t_2)},\label{phiphi}\\ && \langle \varphi_+(t_1)\varphi_-(t_2)\rangle=
i\int\limits_{-\infty}^{\infty} \frac{d\omega}{2\pi} \left(\frac{1}{{C\omega^2/e^2+\cal R}_\omega} \right) e^{-i\omega(t_1-t_2)},\nonumber
\\ && \langle \varphi_-(t_1)\varphi_-(t_2)\rangle=0. \nonumber
\end{eqnarray}

Now we will employ the above results in order to describe the
effect of electron-electron interactions on transport properties
of superconducting contacts.

\section{Coulomb blockade of Andreev reflection}

We start from NS systems. In this case in the absence of interactions
we set $\dot\varphi_{+}=eV$ and trivially recover the standard
result $I=G_AV$. For the current fluctuations $\delta I(t)$ from
Eqs. (\ref{finalS})-(\ref{SINS}) and (\ref{corr}) analogously to \cite{GZ01} we obtain
\begin{equation}
\frac{\langle |\delta I|^2_\omega\rangle}{G_A}
=(1-\beta_A)\omega \coth \frac{\omega}{2T}
+\frac{\beta_A}{2}\sum_{\pm}(\omega \pm 2eV )\coth \frac{\omega
\pm 2eV}{2T}. \label{sn1}
\end{equation}
This equation fully describes current noise in NS structures at energies
well below the superconducting gap. For $eV \gg T,\omega$ Eq. (\ref{sn1})
reduces to the result \cite{dJB}  while in the diffusive regime the correlator  (\ref{sn1})  matches with the semiclassical result \cite{NB}.

Let us now turn on interactions. In this case one should add the
charging term (\ref{Sc}) to the action and account for phase
fluctuations. Proceeding along the same lines as in \cite{GZ01},
for $g_A \gg 1$ or max$(T,eV) \gg E_C=e^2/2C$ we get
\begin{equation}
I=G_AV-2e\beta_A T{\rm Im}\left[ w\Psi\left(1+\frac{w}{2}
\right)-iv\Psi\left(1+\frac{iv}{2} \right) \right]. \label{iv}
\end{equation}
where $\Psi (x)$ is the digamma function, $w=g_AE_C/\pi^2T+iv$ and
$v=2eV/\pi T$.

The last term in Eq. (\ref{iv}) is the interaction correction to
the I-V curve which scales with Andreev Fano factor $\beta_A$ in
exactly the same way as the shot noise. Thus, we arrive
at an important conclusion: {\it interaction correction to Andreev
conductance of NS structures is proportional to the shot noise
power in such structures}. This fundamental relation between
interaction effects and shot noise goes along with that
established earlier for normal conductors \cite{GZ01} extending
it to superconducting systems. In both cases this relation is due
to discrete nature of the charge carriers passing through the
conductor.

Another important observation is that the interaction correction
to Andreev conductance defined in Eq. (\ref{iv}) has exactly the
same functional form as that for normal conductors, cf. Eq. (25)
in \cite{GZ01}. Furthermore, in a special case of diffusive
systems we have $G_N=G_A$, $\beta_N=\beta_A=1/3$ and the only difference between the
interaction corrections to the I-V curve in normal and NS systems
is the charge doubling in the latter case. As a result, the
Coulomb dip on the I-V curve of a diffusive NS system at any given
$T$ is exactly {\it 2 times narrower} than that in the normal
case. We believe that this narrowing effect was detected in normal
wires attached to superconducting electrodes \cite{Bezr}, cf. Fig.
3c in that paper.

\section{Interaction correction to supercurrent}

Let us now turn to the electron-electron interaction correction to the equilibrium Josephson current (\ref{Ichi}). Previously such correction was analyzed in the case of Josephson tunnel barriers in the presence of linear Ohmic dissipation \cite{SZ}. The task at hand is to investigate the interaction correction to the supercurrent in contacts with arbitrary transmission distribution.

In order to evaluate the interaction correction it is necessary to go beyond the Gaussian effective action (\ref{finalsS})-(\ref{SIII}) and to
evaluate the higher order contribution $\sim \varphi^3$. It is easy to observe that the interaction correction to the supercurrent is provided
by the following non-Gaussian terms in the effective action:
\begin{eqnarray}
&& \delta (iS_t)=\int\int\int dt_{1} dt_{2}dt_{3}Y(t_1,t_2,t_3)\varphi_-(t_1) \varphi_+(t_2)\varphi_+(t_3)\nonumber\\ && +\int\int\int dt_{1} dt_{2}dt_{3}Z(t_1,t_2,t_3)\varphi_+(t_1) \varphi_-(t_2)\varphi_-(t_3).
\label{3rd}
\end{eqnarray}
The function $Y(t_1,t_2,t_3)$ can be written as
\begin{eqnarray}
Y(t_1,t_2,t_3)=\int \int \frac{d\omega_1}{2\pi} \frac{d\omega_2}{2\pi}Y(\omega_1,\omega_2) e^{-i\omega_1(t_1-t_2)} e^{-i\omega_2(t_1-t_3)},
\end{eqnarray}
where $Y(\omega_1,\omega_2)=Y(\omega_2,\omega_1)$. The function $Z(t_1,t_2,t_3)$ can be expressed in a similar way.

Adding the non-Gaussian terms (\ref{3rd}) to the action and employing Eq. (\ref{curr}) we arrive at the following expression for the interaction correction
\begin{eqnarray}
\delta I_S(\chi)=ie\int \limits_{-\infty}^{\infty} \frac{d\omega}{2\pi} Y(\omega,-\omega) \langle \varphi_+\varphi_+\rangle_\omega +2ie\int \limits_{-\infty}^{\infty} \frac{d\omega}{2\pi} Z(0,-\omega)\langle \varphi_+\varphi_-\rangle_\omega , \label{zs}
\end{eqnarray}
where the phase-phase correlators are defined in Eq. (\ref{phiphi}).

Let us consider the first term in the right-hand side of Eq. (\ref{zs}).
It is easy to see that in the limit of low temperatures only frequencies $|\omega|>\Delta+\epsilon_n(\chi)$ contribute to the integral in Eq. (\ref{phiphi}) for $\langle \varphi_+\varphi_+\rangle$ while the
contribution from the frequency interval $|\omega|<\Delta+\epsilon_n(\chi)$ vanishes. Furthermore, the leading contribution from the first term in  Eq. (\ref{zs}) is picked up logarithmically from the interval $2\Delta\ll |\omega| \ll 1/R_N C$ where
\begin{equation}
 \langle \varphi_+\varphi_+\rangle_\omega \simeq \frac{e^2 R_N}{|\omega |}
\label{as}
\end{equation}
and the function $Y(\omega,-\omega)$ tends to a frequency independent value.

After a straightforward but tedious calculation in the interesting frequency range  $\omega\gg \Delta$ from Eq. (\ref{St}) one finds
\begin{eqnarray}
Y(\omega,-\omega)=\frac{i\Delta\sin\chi}{4}\sum_n \frac{T_n(1-T_n)(2-T_n\sin^2(\chi/2))}{(1-T_n\sin^2(\chi/2))^{3/2}}
F(\epsilon_n(\chi)).\label{yw}
\end{eqnarray}
This high-frequency term involves the factor $1-T_n$, i.e. it vanishes for fully open conducting channels. Combining Eqs. (\ref{as}), (\ref{yw}) with (\ref{zs}), we arrive at the expression for the supercurrent
\begin{equation}
I(\chi )=I_S(\chi )+\delta I_S(\chi).
\end{equation}
In the limit of low temperatures the interaction correction reads
\begin{eqnarray}
\delta I_S(\chi)=-\frac{e\Delta}{2g_N}\ln\left(\frac{1}{2\Delta R_NC} \right) \sin\chi   \sum_n \frac{ T_n(1-T_n)}{(1-T_n\sin^2(\chi/2))^{3/2}}\left( 2-T_n\sin^2 \frac{\chi}{2}\right), \label{intcor}
\end{eqnarray}
where $g_N=2\pi /(e^2R_N)$ is the dimensionless normal state conductance of the contact. This result is justified as long as the Coulomb correction $\delta I_S(\chi)$ remains much smaller than the non-interacting term
$I_S(\chi )$ (\ref{Ichi}). Typically this condition requires the
dimensionless conductance to be large $g_N\gg\ln (1/2\Delta R_NC)$.

Note that Eq. (\ref{intcor}) was derived only from the first term in Eq.
(\ref{zs}). The second term in this equation involving the function
$Z(0,-\omega)$ and the correlator $\langle \varphi_+\varphi_-\rangle$ can be treated analogously. It turns out to be smaller than that of the first term by the logarithmic factor $\sim\ln (1/2\Delta R_N C)$.

Let us emphasize again an important property of the result (\ref{intcor}): The interaction correction contains the factor $1-T_n$ and, hence, vanishes for fully open barriers. In other words, {\it no Coulomb blockade of the Josephson current is expected in fully transparent superconducting contacts}.

The expression for the interaction correction (\ref{intcor}) can further be specified in the case of diffusive contacts. In the absence of interactions the Josephson current in such contacts follows from (\ref{Ichi}) and takes the form corresponding to the zero-temperature limit of a well known Kulik-Omelyanchuk formula for a short diffusive wire
\begin{equation}
I_S(\chi)=\frac{\pi \Delta}{2 e R_N}\cos\frac{\chi}{2}\ln\frac{1+\sin\frac{\chi}{2}}
{1-\sin\frac{\chi}{2}}.
\end{equation}
Including interactions and averaging (\ref{intcor})
with the bimodal transmission distribution
\begin{equation}
P(T_n)\propto \frac{1}{T_n\sqrt{1-T_n}}.
\label{bimodal}
\end{equation} one finds
\begin{eqnarray}
\delta I_S(\chi)=-\frac{e}{8}\Delta \ln\left( \frac{1}{2\Delta R_N C}\right)\cot (\chi/2) \left[\left(\sin\frac{\chi}{2}+\sin^{-1}\frac{\chi}{2}\right)
\ln\frac{1+\sin(\chi/2)}{1-\sin(\chi/2)}-2\right].
\label{intdif}
\end{eqnarray}

Note that the result (\ref{intcor}) can formally be reproduced if one substitutes $T_n\to T_n+\delta T_n$ into
Eq. (\ref{Ichi}), where
\begin{equation}
\delta T_n=-\frac{2}{g_N}\ln \left(\frac{1}{2\Delta R_NC} \right) T_n(1-T_n),
\label{tcor}
\end{equation}
and then expands the result to the first order in $\delta T_n$.
Interestingly, the same transmission renormalization (\ref{tcor}) follows from the renormalization group (RG) equations \cite{KN,GZ04,BN}
\begin{equation}
\frac{ d T_n}{d L}=-\frac{ T_n(1-T_n)}{\sum_k T_k},\quad L=\ln \left(\frac{1}{\epsilon R_N C} \right)
\end{equation}
derived for {\it normal} conductors. In order to arrive at Eq. (\ref{tcor}) one should just start the RG flow at $\epsilon=1/R_NC$ and stop it at $\epsilon=2\Delta$. Thus, the result (\ref{intcor}) can be interpreted in a very simple manner: Coulomb interaction provides high frequency renormalization $T_n+\delta T_n$ (\ref{tcor}) of the barrier transmissions which should be substituted into the classical expression for the supercurrent (\ref{Ichi}).
It should be stressed, however, that the last step would by no means appear obvious without our rigorous derivation since the Coulomb correction to the Josephson current originates from the term $\sim \varphi_-\varphi_+^2$ in the effective action which is, of course, totally absent in the normal case.

\section{Summary}

In this paper we derived a general  expression for the effective action of superconducting contacts with arbitrary transmissions of
conducting channels. In the case of NS systems we described the interplay between Coulomb blockade and Andreev reflection and demonstrated a direct
relation between shot noise and interaction effects in these structures. The fundamental physical reason behind this relation lies in discrete nature of the charge carriers -- electrons and Cooper
pairs -- passing through NS interfaces. Our results allow to explain recent experimental findings \cite{Bezr}.

Superconducting contacts with arbitrary channel transmissions show qualitatively new features
as compared to the case of Josephson tunnel barriers \cite{SZ}.
The main physical reason for such differences is the presence of subgap Andreev bound states inside the system.
Our results for the interaction correction might explain a rapid change between superconducting and insulating behavior recently observed  \cite{Bezr2} in
comparatively short metallic wires with resistances close to
the quantum resistance unit $\sim 6.5$ K$\Omega$ in-between two bulk
superconductors. Previously it was already
argued \cite{AGZ} that such a superconductor-to-insulator crossover
can be due to Coulomb effects. Our present results provide further
quantitative arguments in favor of this conclusion.

\vspace{0.3cm}

 \centerline{\bf
Acknowledgment}

\vspace{0.3cm}

This work was supported in part by RFBR grant 09-02-00886.


\section*{References}
\begin{thebibliography}{50}
\bibitem{And} Andreev A.F. 1964 {\it Sov. Phys. JETP} {\bf 19} 1228.
\bibitem{BTK} Blonder G.E., Tinkham M., and Klapwijk T.M. 1964 {\it Phys. Rev. B} \textbf{25} 4515.
\bibitem{SNSK} Kulik I.O.1970 {\it Sov. Phys. JETP} {\bf 30} 944.
\bibitem{SNSI} Ishii C. 1970 {\it Progr. Theor. Phys.} {\bf 44} 1525.
\bibitem{SNSB} Beenakker C.W.J. 1991 {\it Phys. Rev. Lett.} {\bf 67} 3836.
\bibitem{SNSGZ} Galaktionov A.V. and Zaikin A.D. 2002 {\it Phys. Rev. B} {\bf 65} 184507.
\bibitem{SZ} Sch\"on G. and Zaikin A.D. {\it Phys. Rep.} 1990 {\bf 198} 237.
\bibitem{GZ01} Golubev D.S. and Zaikin A.D. 2001 {\it Phys. Rev. Lett.} {\bf 86} 4887.
\bibitem{GGZ03} Galaktionov A.V., Golubev D.S., and Zaikin A.D. 2003 {\it Phys. Rev. B} {\bf 68} 085317; {\it ibid.} {\bf 68} 235333.
\bibitem{KN} Kindermann M. and Nazarov Yu.V. 2003 {\it Phys. Rev. Lett.} {\bf 91} 136802.
\bibitem{GZ04} Golubev D.S. and Zaikin A.D. 2004 {\it Phys. Rev. B} {\bf 69} 075318.
\bibitem{BN} Bagrets D.A. and Nazarov Yu.V. 2005 {\it Phys. Rev. Lett.} {\bf 94} 056801.
\bibitem{Z} Zaikin A.D. 1994 {\it Physica B} \textbf{203} 255.
\bibitem{SN} Snyman I. and Nazarov Yu.V. 2008 {\it Phys. Rev. B}
{\bf 77} 165118.
\bibitem{GZ09} Galaktionov A.V. and Zaikin A.D. 2009 {\it Phys. Rev. B} {\bf 80} 174527.
\bibitem{KO} Kulik I.O. and Omel'yanchuk A.N. 1978 {\it Sov. J. Low Temp. Phys.} {\bf 4} 142.
\bibitem{GZ2} Golubev D.S. and Zaikin A.D. 1999 {\it Phys. Rev} B {\bf 59} 9195.
\bibitem{dJB} De Jong M.J.M. and
Beenakker C.W.J. 1994 {\it Phys. Rev. B} {\bf 49} 16070.
\bibitem{NB} Nagaev K.E. and B\"uttiker M. 2001 {\it Phys. Rev. B}
{\bf 63} 081301.
\bibitem{Bezr} Bollinger A.T., Rogachev A., and Bezryadin A. 2006 {\it Europhys. Lett.} {\bf 76} 505.
\bibitem{Bezr2} Bollinger A.T., Dinsmore III R.C., Rogachev A., and Bezryadin A. 2008 {\it Phys. Rev. Lett.} {\bf 101} 227003.
\bibitem{AGZ} Arutyunov K.Yu., Golubev D.S., and Zaikin A.D.
2008 {\it Phys. Rep.} {\bf 464} 1.
\end{thebibliography}
\end{document}